\documentclass{article}
\usepackage{ijcai26}

\usepackage{times}
\usepackage{soul}
\usepackage{url}
\usepackage[hidelinks]{hyperref}
\usepackage[utf8]{inputenc}
\usepackage[small]{caption}
\usepackage{graphicx}
\usepackage{amsmath}
\usepackage{amsthm}
\usepackage{booktabs}
\usepackage{algorithm}
\usepackage{algorithmic}
\usepackage[switch]{lineno}

\usepackage{xcolor}
\usepackage{tabularx}
\usepackage{ragged2e}
\usepackage{multirow}

\title{ConceptCaps: a Distilled Concept Dataset for Interpretability in Music Models}

\author{
Bruno Sienkiewicz$^1$
\and
Łukasz Neumann$^1$\And
Mateusz Modrzejewski$^1$\\
\affiliations
$^1$Institute of Computer Science, Warsaw University of Technology\\
\emails
bruno.sienkiewicz@gmail.com,
\{lukasz.neumann, mateusz.modrzejewski\}@pw.edu.pl,
}

\begin{document}

\maketitle

\begin{abstract}
Concept-based interpretability methods like TCAV require clean, well-separated positive and negative examples for each concept. Existing music datasets lack this structure: tags are sparse, noisy, or ill-defined. We introduce ConceptCaps, a dataset of 21k music-caption-tags triplets with explicit labels from a 200-attribute taxonomy. Our pipeline separates semantic modeling from text generation: a VAE learns plausible attribute co-occurrence patterns, a fine-tuned LLM converts attribute lists into professional descriptions, and MusicGen synthesizes corresponding audio. This separation improves coherence and controllability over end-to-end approaches. We validate the dataset through audio-text alignment (CLAP), linguistic quality metrics (BERTScore, MAUVE), and TCAV analysis confirming that concept probes recover musically meaningful patterns. Dataset and code are available \href{https://anonymized-for-the-blind-review.com}{online}.
\end{abstract}

\section{Introduction}

Work on interpretability and explainability in music information retrieval has increasingly turned to “concept-level” analyses, such as Testing with Concept Activation Vectors (TCAV \cite{kim2018interpretability}), which offers a promising framework for interpreting these models by identifying and measuring sensitivity to high-level musical aspects like instrumentation, genre, and mood \cite{foscarin2022concept,gebhardt2025beyond,afchar2022learning}. In practice, however, these concepts are usually inferred from free-form captions, weak tags, or proxy tasks that do not provide explicit, validated concept labels. This leads to a gap between the theoretical object of analysis (a well-defined musical concept) and the empirical evidence available in existing datasets, which are dominated by abstract, overlapping, and sometimes contradictory descriptions.

This gap has concrete consequences. First, it is difficult to construct positive and negative example sets that isolate a concept from closely related phenomena, which undermines the design of controlled experiments. Second, comparisons across models and studies are hard to interpret, because each work often relies on ad hoc concept definitions and bespoke data slices. Third, it is unclear how much of a reported “concept sensitivity” reflects genuine semantic structure in the model, and how much is an artefact of annotation noise or dataset bias.

To address these limitations, we propose a novel dataset generation pipeline based on \emph{separation of concerns}. Similar methods often rely on asking a single LLM to both generate plausible musical attributes and write fluent descriptions. We decompose the task into two specialized stages. First, a Variational Autoencoder (VAE) \cite{kingma2014autoencoding} learns the distribution of musical attribute co-occurrence from curated source data, ensuring that sampled combinations are statistically plausible and musically coherent. Second, a fine-tuned language model translates these controlled attribute lists into professional, context-aware music descriptions. This separation reduces hallucination, improves controllability, enables independent optimization of semantic consistency and linguistic quality, and yields a dataset explicitly designed to support concept-based interpretability analyses.

\subsection{Main Contributions}

The primary contributions of this work are:

\begin{enumerate}
\item We introduce ConceptCaps, a large-scale, copyright-free dataset of 21k music-description pairs with rich, validated concept labels and matched counterexamples, specifically designed for concept-based explainability research.

\item We present a novel two-stage generative pipeline that separates VAE-based semantic consistency from fine-tuned LLM linguistic quality, enabling controlled, high-quality music dataset generation while maintaining efficiency through local, reproducible processing.

\item We demonstrate quality and utility using linguistic metrics (BLEU, ROUGE, BERTScore, MAUVE), audio-text alignment (CLAP scores), concept-specific metrics, and downstream TCAV analysis demonstrating both dataset quality and interpretability.

\item We show our approach achieves competitive results with significantly lower computational cost than API-based methods \cite{ouyang2022training}, and provides fine-grained control over dataset characteristics through VAE latent space sampling and conditioning.
\end{enumerate}

\section{Related Work}
\label{sec:related}

\subsection{Concept-Based Explainability in Deep Learning}

Concept-based methods such as Testing with Concept Activation Vectors (TCAV)~\cite{kim2018interpretability} have become increasingly popular for interpreting neural networks across vision and audio domains. TCAV enables quantitative measurement of model sensitivity to user-defined concepts by extracting directions in neural network activations that correspond to those concepts. The core insight is that understanding decision-making at the concept level---rather than at the input or neuron level---can provide more intuitive and actionable insights.

However, TCAV's effectiveness depends critically on having clean, well-separated positive and negative concept examples. When concept examples are noisy, sparse, or poorly defined, TCAV scores become unreliable and difficult to interpret. This bottleneck has limited the adoption of concept-based analysis in music research, where high-quality concept datasets are scarce. Our work directly addresses this limitation by providing a systematic, reproducible method to construct such datasets. Related interpretability approaches in audio and music domains have demonstrated the potential of concept-level analysis~\cite{elizalde2023clap,zhou2022interpretability,gebhardt2025beyond,foscarin2022concept,afchar2022learning}.

\subsection{Music Dataset Augmentation and Synthesis}

Recent work on music captioning has explored using large language models to augment existing music datasets. Approaches such as LP-MusicCaps~\cite{doh2023lp} and WavCaps~\cite{mei2024wavcaps} generate or refine captions using language models, demonstrating the potential of LLM-based data augmentation. While these approaches effectively scale data production, they face two key limitations in the interpretability context:

\begin{enumerate}
\item \textbf{Lack of Controllability:} Generated data follows the biases of the LLM's training distribution rather than explicitly controlling concept patterns needed for concept analysis.

\item \textbf{Limited Semantic Validation:} These methods often rely on existing datasets that can be noisy and lack systematic representation of specific concepts. Without validation of semantic coherence, it is unclear whether generated attributes form meaningful, interpretable patterns.
\end{enumerate}

\section{Method}
\label{sec:method}

Our architecture draws inspiration from successful multi-stage generation approaches in computer vision and other domains, like StackGAN~\cite{han2017stackgan} or CLAP~\cite{elizalde2023clap}. Our pipeline adopts this principle of \textbf{separation of concerns}: the VAE acts as a ``Stage-I'' generator sketching the semantic skeleton (musical attributes and their plausible combinations), while the fine-tuned LLM acts as a ``Stage-II'' refiner, constructing descriptions with linguistic detail, professional vocabulary, and contextual nuance.

\subsection{Pipeline Overview}

At a high level, source-derived attributes are first distilled into a consistent taxonomy with representative examples, then modeled as co-occurrence patterns by the VAE. The fine-tuned LLM translates these attribute lists into natural language descriptions, and MusicGen synthesizes corresponding audio. This ensures that each synthetic sample is explicitly paired with concept labels drawn from the learned distribution and representative audio realizations.

\begin{figure}[t]
\centering
\includegraphics[width=1.0\linewidth]{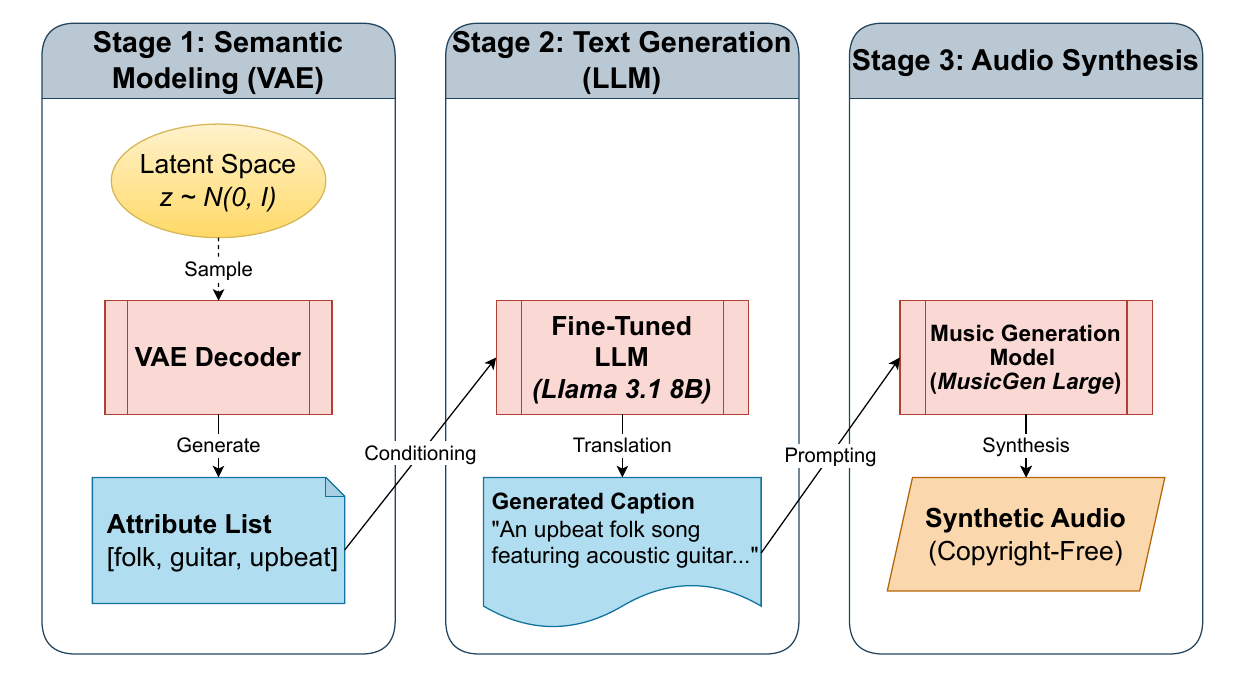}
\caption{Overview of the three-stage dataset generation pipeline. \textbf{Stage 1 (Semantic Modeling via VAE):} A Variational Autoencoder samples from the latent space $z \sim \mathcal{N}(0, I)$ and decodes to generate coherent attribute lists (e.g., ``folk, acoustic guitar, upbeat''). \textbf{Stage 2 (Text Generation via Fine-Tuned LLM):} The fine-tuned language model (Llama 3.1 8B) receives the attribute list and generates a professional, descriptive music caption capturing the semantic content. \textbf{Stage 3 (Audio Synthesis):} MusicGen synthesizes copyright-free audio from the generated description, completing the dataset sample with well-aligned audio-text pairs and explicit concept labels suitable for interpretability analysis.}
\label{fig:pipeline}
\end{figure}

\subsection{Dataset Distillation \& Concept Taxonomy}

To obtain a high-quality, concept-dense dataset suitable for rigorous interpretability analysis, we must first define a meaningful concept taxonomy and source high-quality training data. The MusicCaps dataset \cite{agostinelli2023musiclm} contains 5,521 human-annotated descriptions with associated tags, providing a valuable starting point. However, MusicCaps suffers from sparse and noisy categories. These issues undermine the quality of downstream semantic models. We therefore apply a distillation methodology to ensure our VAE and LLM components receive clean, representative data.

The distillation procedure consists of three steps:

\begin{enumerate}
\item \textbf{Tag Extraction and Organization:} Identify all unique tags in MusicCaps and group them into coherent semantic categories such as instruments (acoustic guitar, piano, drums), genres (folk, electronic, classical), moods (upbeat, melancholic, energetic), and tempo characteristics. This hierarchical organization enables targeted concept analysis.

\item \textbf{Representative Subset Selection:} Select samples that cover aspects from previously defined taxonomy, excluding outliers and samples with sparse or unclear tags. Prioritize samples with multiple tags from different categories to ensure diversity.

\item \textbf{Aspect Quality Refinement:} Extract only samples with tags from at least 3 semantic categories, clear and grammatically sound descriptions, and coherent attribute combinations.
\end{enumerate}

By distilling MusicCaps from 5,521 to 1,890 high-quality samples, we ensure the VAE trains on representative, high-confidence examples rather than noisy or ambiguous data. This smaller, curated subset also yields a clean categorized taxonomy essential for concept-based explanation techniques. The resulting 1,890 pairs of attribute lists and professional captions serve as training data for both VAE co-occurrence modeling and LLM fine-tuning.

\subsection{Semantic Modeling via VAE}

The core motivation behind semantic modelling with VAE is to generate a rich source of \emph{coherent} attribute combinations. This is particularly important in concept-based research, where specific dataset characteristics must be modelled with precision. A naive approach of random attribute sampling reduces controllability and introduces contradictory combinations (e.g., ``quiet death metal''), which can negatively affect downstream analysis by introducing confusion in the studied models.

\begin{figure}[t]
    \centering
    \includegraphics[width=0.8\linewidth]{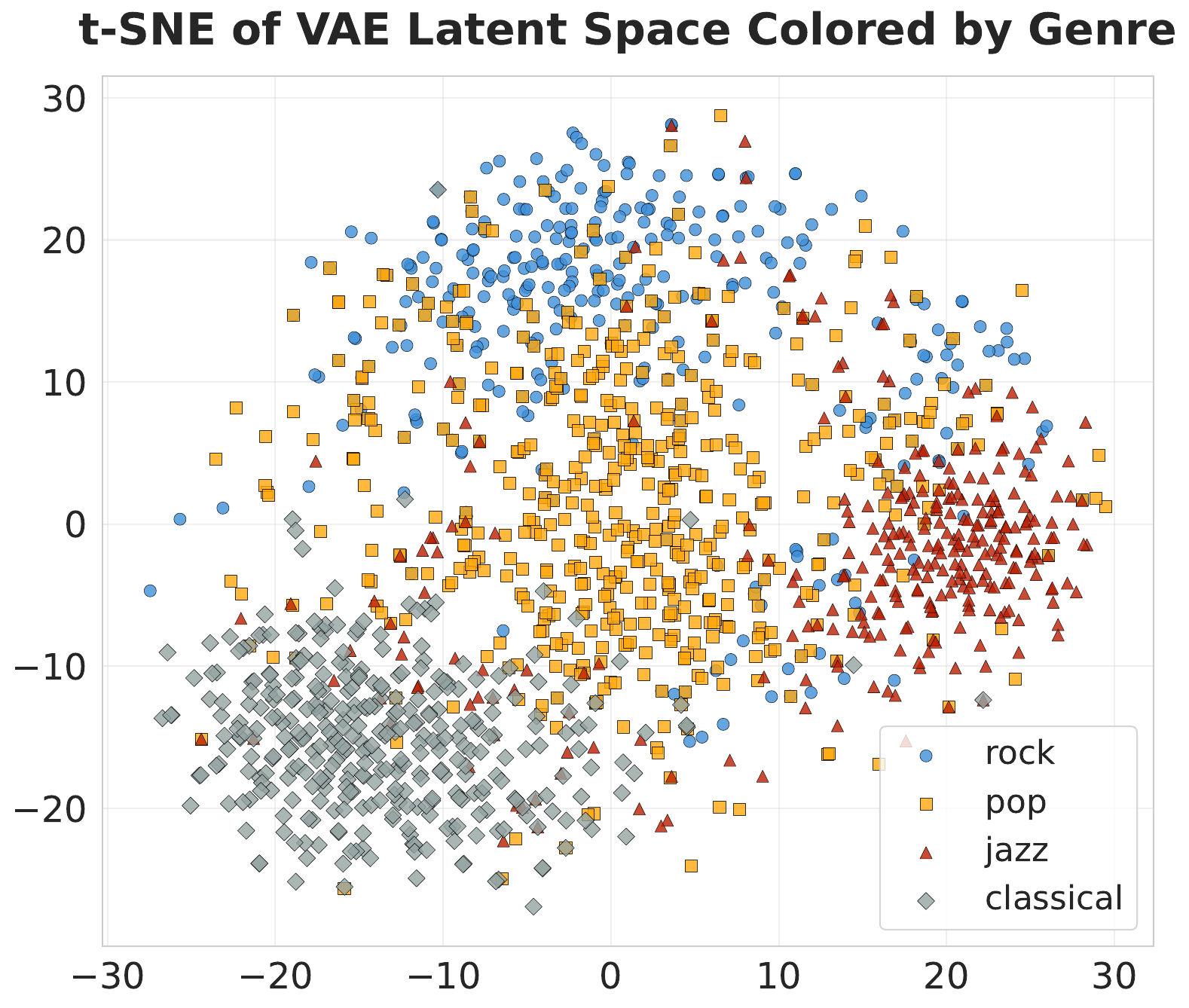}
    \caption{Visualisation of VAE latent space. Encoder successfully learns to map music genres to latent space. We can also notice music genres "blending" together, especially in "rock" and "pop" genres, where conceptual overlap in source dataset is substantial.}
    \label{fig:vae_latent_space}
\end{figure}

We use standard VAE architecture that is trained on multi-hot encoded attribute vectors representing the presence or absence of each musical aspect in our taxonomy. Let $x \in \{0,1\}^D$ be such a binary vector, where $D$ is the number of unique attributes. The model learns to reconstruct such vector by sampling the latent space. This latent space is visualised in Figure \ref{fig:vae_latent_space} to show separation of different musical aspects, while grouping similar characteristics together.

\subsubsection{Inference: Coherent Attribute Vector Generation}

By learning the \emph{distribution} of co-occurrence patterns rather than relying only on examples in the source dataset, the VAE enables generation of new combinations that are statistically plausible while avoiding noise and bias. This is more powerful than either random sampling or simple template-based approaches. It supports both conditional and unconditional sampling by either giving a partial attribute vector $x_{\text{seed}}$ specifying desired properties, or by sampling a latent vector $z \sim \mathcal{N}(0, I)$ and passing it to Decoder.

\subsection{Controlled Text Generation via Fine-Tuned LLM}

We found that direct prompting of pre-trained Large Language Models (LLMs) with raw attribute lists frequently yields suboptimal results, characterized by hallucinations or excessive verbosity that dilutes semantic precision.

Our solution is to fine-tune a pre-trained, instruction-tuned language model (Llama 3.1 8B)~\cite{meta2024llama31} on human-made captions found in MusicCaps dataset. By utilizing Quantized Low-Rank Adaptation (QLoRA)~\cite{dettmers2023qlora}, this approach is significantly more efficient than full fine-tuning, while preserving general linguistic knowledge and specializing the model for music description. QLoRA reduces memory requirements through 4-bit quantization while maintaining performance via rank-32 low-rank adapters.

The model is fine-tuned on pairs $(A, C)$ from our distilled dataset, where $A$ is an attribute list and $C$ is a ground-truth professional caption. By fine-tuning on attribute-caption pairs, the model learns to prioritize semantic density and directly incorporate provided attributes into descriptions. The conditioning shifts the learning objective from ``generate plausible song description'' to ``write professional, accurate descriptions of given attributes''. This is a key distinction that improves both quality and controllability.

Our ablation studies (Section \ref{sec:ablation}) demonstrate that this fine-tuned LLM approach significantly outperforms zero-shot and base-model approaches.

\subsection{Audio Synthesis}

The final stage uses MusicGen~\cite{copet2023simple} to synthesize audio conditioned on the LLM-generated descriptions. MusicGen was trained exclusively on Meta-owned and licensed music, avoiding legal uncertainties of models trained on scraped material \cite{batlleroca2025musgo}. Its weights are released under CC-BY-NC 4.0, restricting use to non-commercial research. The resulting dataset is therefore both fully suitable for academic purposes and reproducible within these terms.

\subsection{Implementation}

% TODO link do repo
% TODO deanonymize
Source code, datasets and pre-trained models are publicly available online for reproducibility purposes\footnote{\url{https://github.com/BrunoSienkiewicz/ConceptCaps}}. We are using $\beta$-VAE~\cite{higgins2017betavae} variant of variational auto-encoder with additional $\beta$ parameter, which controls KL divergence loss. From our experiments we concluded that $\beta=0.25$ yielded the best overall results. It suggests that minimizing reconstruction loss is more important in our case than mapping latent space to normal distribution. Higher $\beta$ values resulted in model collapse suggested by VAE predicting only the few most popular attribute combinations. VAE is implemented with a standard fully-connected architecture: encoder with $D$-512 hidden units, where $D$ is denotes our aspect taxonomy dimension, mapping to a 128-dimensional latent space, and a symmetric decoder. We use Adam optimizer \cite{kingma2014adam} with learning rate $3 \times 10^{-4}$ and train for 110 epochs on the 1,890 distilled samples. 

For LLM fine-tuning, we use the Hugging Face Transformers \cite{wolf2020transformers} library with QLoRA configuration: rank 32, lora-alpha 8, dropout 0.26. We fine-tune for 5 epochs with batch size 4 using the distilled attribute-caption pairs. 

Audio synthesis uses MusicGen with increased guidance scale of 3.3 for more concept alignment of audio inference resulting in 30-second clip for each caption. Detailed evaluation of our implementation with explanation of each metric on each stage of the pipeline can be found in Table \ref{tab:arch_params}.

\begin{table*}[t]
\centering
\caption{Metrics for Best-Performing Hyperparameters Across Stages.}
\label{tab:arch_params}
% \begin{tabular}{llcl}
\begin{tabularx}{\textwidth}{@{} p{1.2cm}ll >{\RaggedRight}X @{}}
\toprule
\textbf{Pipeline Stage} & \textbf{Metric} & \textbf{Value} & \textbf{Interpretation} \\ \midrule
\multirow{7}{*}{\parbox{1.2cm}{\textbf{Stage 1: VAE}}} & 
BCE $\downarrow$ &
105.21 &
Binary Cross-Entropy loss of the reconstructed vectors vs. ground-truth attribute vectors. \\
&
Jaccard $\uparrow$ & 
0.716 & 
Intersection-over-union of reconstructed vs. ground-truth attribute vectors. \\
& 
Hamming $\downarrow$ & 
0.005 & 
Mean element-wise prediction error on individual attributes. \\
& 
Diversity \% $\uparrow$ & 
96.03\% & 
Proportion of unique attribute vectors generated from 1K samples. \\
& 
Cosine Similarity $\uparrow$ & 
0.808 & 
Cosine distance between learned co-occurrence patterns and ground-truth MusicCaps patterns. \\ \midrule
\multirow{8}{*}{\parbox{1.2cm}{\textbf{Stage 2: LLM}}} & 
BERTScore-F1 \cite{zhang2020bertscore} $\uparrow$ & 
0.8988 & 
Contextual semantic similarity using BERT embeddings between generated and ground-truth captions. \\
& 
BLEU \cite{papineni2002bleu} $\uparrow$ & 
0.2119 & 
Unigram, bigram, trigram, and 4-gram overlap with reference captions.  \\
& 
ROUGE-L \cite{lin2004rouge} $\uparrow$ & 
0.3906 & 
Longest common subsequence F-score between generated and reference captions. \\
& 
MAUVE \cite{pillutla2021mauve} $\uparrow$ & 
0.9503 & 
Distribution divergence metric between generated and human-written caption distributions. \\ \midrule
\multirow{4}{*}{\parbox{1.2cm}{\textbf{Stage 3: Audio}}} & 
CLAP \cite{elizalde2023clap} $\uparrow$ & 
0.5563 &
Similarity between textual description and audio representation calculated using pre-trained CLAP model. \\
& 
FAD \cite{kilgour2019frechet} $\downarrow$ & 
0.5135 & 
Fréchet Audio Distance, quantifying divergence between real and generated audio distributions using CLAP embeddings. \\ \bottomrule
% \end{tabular}
\end{tabularx}
\end{table*}

\begin{table*}[t] 
\centering 
\small 
\caption{Comparison of Music-Text and Audio-Language Datasets} 
\label{tab:dataset_comparison} 
\begin{tabularx}{\textwidth}{p{2cm}lcp{2.5cm}XXXX} 
\toprule
\textbf{Type} & \textbf{Dataset} & \textbf{\# Samples} & \textbf{\# Unique Aspects} & \textbf{Caption Available} & \textbf{Avg. Length (words)} & \textbf{Total Duration (h)} \\ 
\midrule
\multirow{5}{*}{\parbox{2cm}{\textit{Human-Annotated}}} & AudioSet \cite{gemmeke2017audioset} & $\sim$2,100,000 & 527 classes & No & N/A & $\sim$5,800 \\ 
& AudioCaps \cite{kim2019audiocaps} & 51,308 & N/A & Yes & $\sim$9 & 142.5 \\ 
& MusicCaps \cite{agostinelli2023musiclm} & 5,521 & $\sim$500 & Yes & $\sim$48 & 15.3 \\ 
& GTZAN \cite{tzanetakis2002musical} & 1,000 & 10 genres & No & N/A & 8.3 \\
& SongDescriber \cite{manco2023songdescriber} & 1,102 & N/A & Yes & $\sim$22 & 23.5 \\ 
\midrule
\multirow{4}{*}{\parbox{2cm}{\textit{LLM/Synthetic-Generated}}} & LP-MusicCaps \cite{doh2023lp} & 512,165 & $\sim$1,000 & Yes & $\sim$10 & 1,422 \\ 
& WavCaps \cite{mei2024wavcaps} & 403,050 & 527 classes & Yes & $\sim$8 & 7,563 \\ 
& Sound-VECaps \cite{yuan2024soundvecaps} & 1,660,000 & 527 classes & Yes & $\sim$30 & 4,611 \\ 
& \textbf{ConceptCaps (Ours)} & \textbf{21,433} & \textbf{200} & \textbf{Yes} & \textbf{$\sim$31} & \textbf{178} \\ 
\bottomrule
\end{tabularx} 
\end{table*}

\section{Dataset}
\label{sec:dataset}

\subsection{Dataset Overview}

We present a large-scale dataset composed of curated attributes representing meaningful musical concepts suitable for interpretability analysis. Our dataset significantly exceeds existing concept-labeled music resources in both scale and semantic quality.

\begin{figure}[t]
\centering
\includegraphics[width=0.95\linewidth]{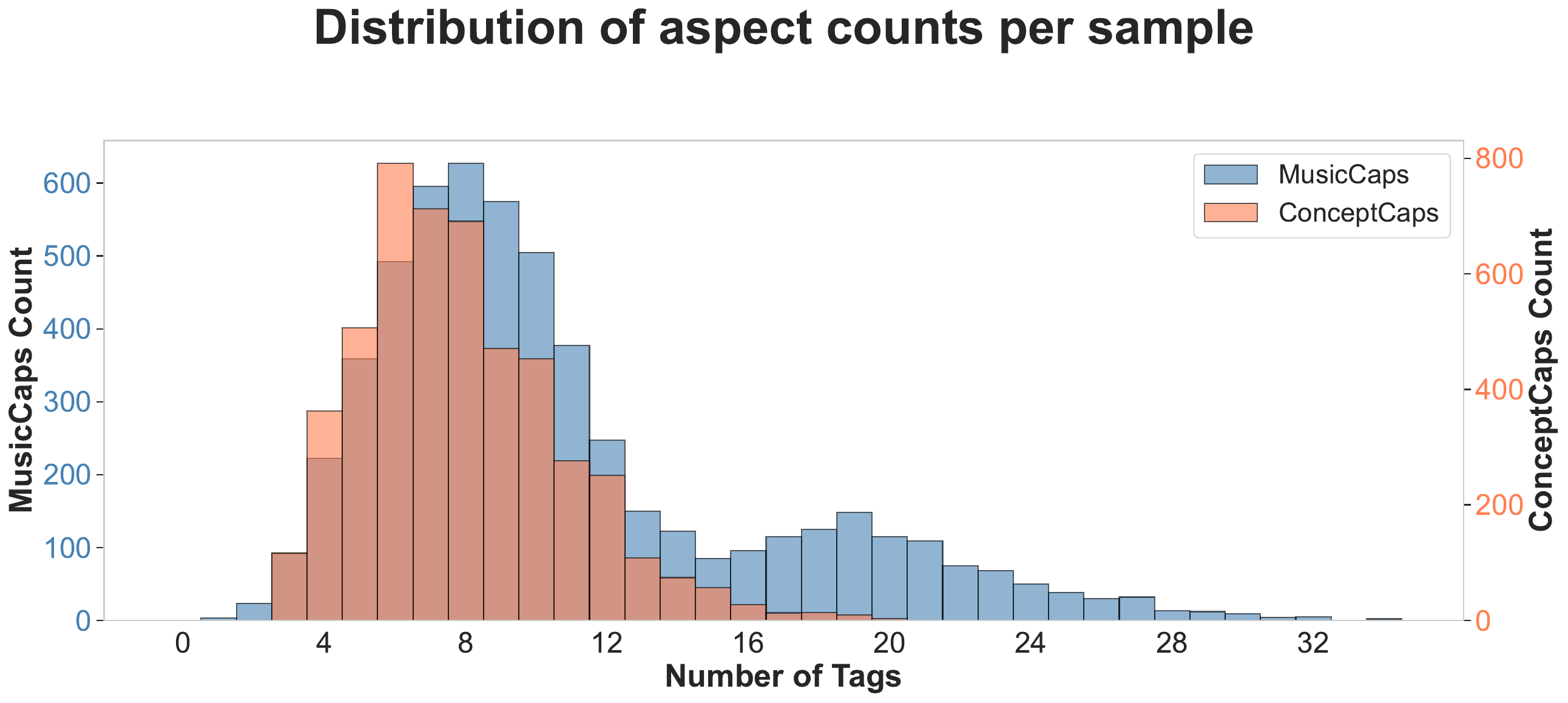}
\caption{Comparison of per-sample aspect count distribution in MusicCaps (blue) versus our distilled dataset (orange). Lack of long tail shows improvement over sparse or redundant tags in original dataset.}
\label{fig:tag_counts}
\end{figure}

Our dataset demonstrates a marked improvement over MusicCaps in terms of concept representation quality and consistency. As seen in Figure \ref{fig:tag_counts}, we substantially reduce the long-tail distribution present in MusicCaps, where sparse and redundant tags complicate down-stream interpretability analysis. By deliberately limiting the number and increasing the quality of concepts per sample, generative models can represent them more faithfully in the final output.

\subsection{Concept Distribution Analysis}

\begin{figure}[t]
\centering
\includegraphics[width=0.9\linewidth]{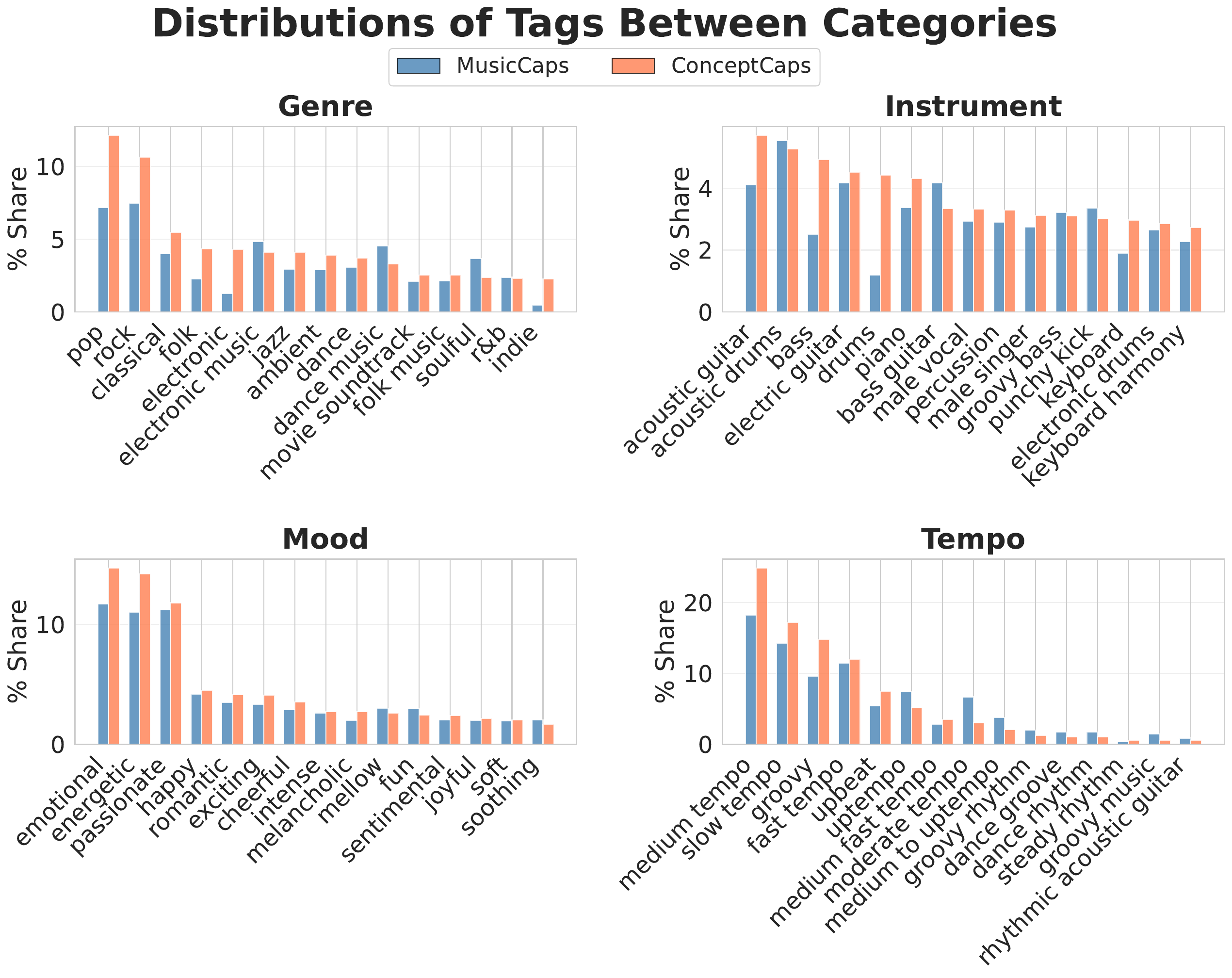}
\caption{Individual tag frequency distributions comparing MusicCaps (blue) and our distilled dataset (orange). The distilled dataset preserves source dataset characteristics, resulting in real-world data distribution.}
\label{fig:individual_tag_distribution}
\end{figure}

The distillation and VAE-based generation process preserves correlation with the original MusicCaps distribution while introducing important improvements in meaningful concept density seen in Figure \ref{fig:individual_tag_distribution}. By employing the VAE to model co-occurrence patterns, we can efficiently represent real-world attribute relationships while maintaining access to vast new, coherent attribute combinations. This is superior to either using only original data (which has limited size) or random sampling (which lacks semantic coherence).

\section{Experiments}
\label{sec:experiments}

To evaluate our distilled concept dataset pipeline, we design experiments across three dimensions: 
\begin{itemize}
    \item In Section \ref{sec:cross-dataset} we access dataset quality relative to existing benchmarks.
    \item In Section \ref{sec:ablation} we validate improvements of individual pipeline stages.
    \item In Section \ref{sec:tcav} we test downstream task performance measuring interpretability utility.
\end{itemize}

Our experiments address a fundamental challenge in dataset curation for concept-based analysis: ensuring that high-quality, semantically coherent attributes map correctly to audio realizations. This requires validation of semantic consistency (VAE stage), caption quality (LLM stage), and audio-text alignment (MusicGen stage). Ablation studies are critical because they isolate the contribution of our two-stage architecture relative to end-to-end approaches, demonstrating that separation of concerns yields measurable improvements in both quality metrics and downstream interpretability performance.

\subsection{Evaluation}

VAE Stage validates that the Variational Autoencoder successfully learns valid musical attribute combinations. As shown in Table \ref{tab:arch_params}, high Jaccard and low Hamming loss confirm accurate reconstruction of individual attributes. High Diversity shows the model generates diverse combinations rather than sticking to the most popular combinations, critical for creating a larger dataset and avoiding model collapse. Cosine similarity ensures learned patterns reflect real musical semantics.

In the LLM Stage, fine-tuned language model achieves robust semantic quality indicated by high BertScore \cite{zhang2020bertscore}, BLEU \cite{papineni2002bleu} and ROUGE-L \cite{lin2004rouge}, confirming significant overlap between reference and ground-truth samples. MAUVE \cite{pillutla2021mauve} validates that overall output characteristics (vocabulary, style, length distribution) match human-written captions, confirming successful adaptation of the fine-tuned model to the music description domain.

In the final stage to validate that the generated descriptions effectively guide the audio synthesis process, we evaluate the semantic alignment between the synthesized audio and its corresponding text using CLAP (Contrastive Language-Audio Pretraining) scores \cite{elizalde2023clap}. CLAP provides a robust metric for semantic similarity that aligns well with human perception of relevance. While recent research \cite{okamoto2025humanclap} suggests that CLAP’s correlation with subjective human judgment can vary, it remains a critical tool for objective, large-scale evaluation where manual annotation is infeasible. Additionally, we measure the Fréchet Audio Distance (FAD) to assess the distribution gap between our generated audio and real-world music. Following the research of \cite{gui2023}, we utilize CLAP-based audio embeddings, as they capture deeper semantic and melodic features that more closely align with human perceptual judgments of "musicality" and "audio quality". We employ the GTZAN dataset \cite{tzanetakis2002musical} as a refrence, providing a benchmark for melodic and rhythmic consistency.

\subsection{Cross-dataset Audio-Text Alignment Quality Comparison}
\label{sec:cross-dataset}

We benchmark our distilled dataset against two primary baselines: \textbf{MusicCaps}, which serves as our human-annotated ground-truth source, and \textbf{LP-MusicCaps}, a large-scale synthetic corpus. While MusicCaps provides a high-quality reference point for assessing the baseline of human-level description, LP-MusicCaps offers a direct methodological comparison. Specifically, both LP-MusicCaps and our work utilize the MusicCaps metadata as a seed; however, LP-MusicCaps lacks the multi-stage taxonomy distillation and targeted LLM fine-tuning that characterize our approach, allowing us to isolate the impact of these specific architectural contributions.

\begin{figure}[t]
\centering
\includegraphics[width=1\linewidth]{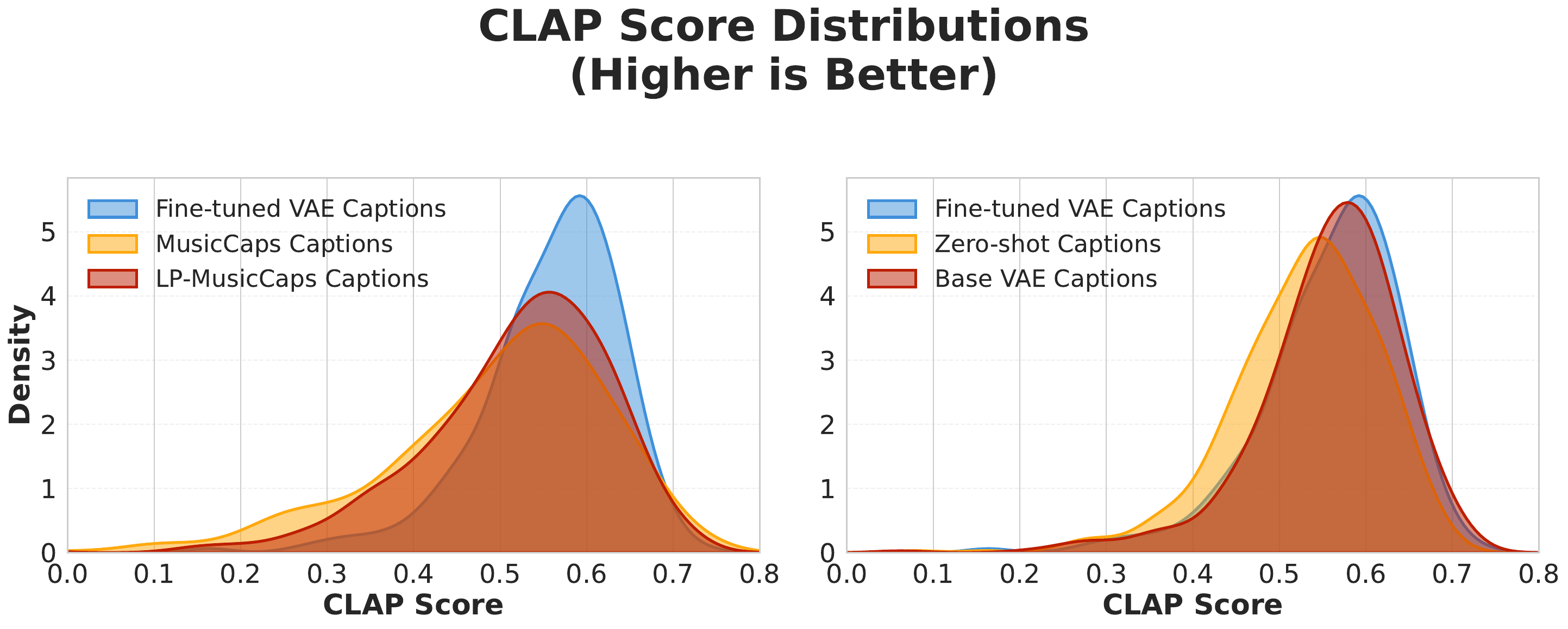}
\caption{CLAP score distributions comparing our distilled dataset against MusicCaps and LP-MusicCaps (left), alongside an ablation study of LLM configurations (right). Higher scores denote superior semantic alignment between audio and text.}
\label{fig:clap_distribution}
\end{figure}

As illustrated in Figure \ref{fig:clap_distribution}, our dataset demonstrates an improvement in audio-text alignment compared to existing synthetic benchmarks. While the human-authored MusicCaps dataset exhibits a notable amount of high-performing samples, it also displays a significantly higher density of samples with near-zero alignment, likely due to noisy or overly abstract human captions. In contrast, our dataset yield a more concentrated distribution with a higher mean score than the LP-MusicCaps baseline, suggesting that our distillation process produces more representative and semantically consistent descriptions for generative models.

\subsection{Ablation Study}
\label{sec:ablation}

We conducted an ablation study to quantify the performance gain provided by the fine-tuning stage of our text generation pipeline. Specifically, we compared our fine-tuned model against the base LLM (Base VAE Captions) and a zero-shot inference baseline. The latter represents a scenario where the LLM generates descriptions without attribute conditioning, serving as a baseline for the need of our VAE-based attribute generation.

As shown in the right-hand panel of Figure \ref{fig:clap_distribution}, the fine-tuned LLM achieves a superior distribution shift compared to both the base model and the zero-shot baseline. The poor performance of the zero-shot model highlights the difficulty of obtaining representative audio samples without explicit attribute control. Crucially, the improvement of the fine-tuned model over the base VAE model confirms that fine-tuning effectively optimizes for \textit{semantic density}—teaching the model to use precise, professional terminology and eliminate verbose filler that can dilute the conditioning signal during audio synthesis.

\subsection{TCAV Analysis: Concept Separability in Music Classifiers}
\label{sec:tcav}

To demonstrate the practical utility of our distilled dataset for concept-based interpretability, we performed Testing with Concept Activation Vectors (TCAV) analysis on a music genre classification task. TCAV quantifies the extent to which a model's internal representations align with high-level, user-defined concepts by measuring the sensitivity of predictions to concept-aligned directions in the latent space.

\subsubsection{Experimental Setup}

We evaluated our distilled dataset by training a standard Convolutional Neural Network (CNN) on the GTZAN \cite{tzanetakis2002musical} benchmark, a foundational corpus for music genre classification. The model attained a validation accuracy of 80\%. While specialized architectures may yield higher absolute performance, this baseline is sufficient to demonstrate that the network has developed discriminative acoustic representations suitable for concept-based probing.

Using our distilled resource, we constructed balanced positive and negative example sets for TCAV analysis. To ensure alignment with the spatial-temporal feature hierarchies learned by the CNN, we focused on "low-to-mid-level" acoustic concepts—specifically tempo and instrumentation—which possess distinct spectral signatures. This approach allows us to verify whether the classifier’s decision-making process is grounded in semantically meaningful musical attributes.

\subsubsection{Results}

\begin{figure}[t]
    \centering
    \includegraphics[width=1\linewidth]{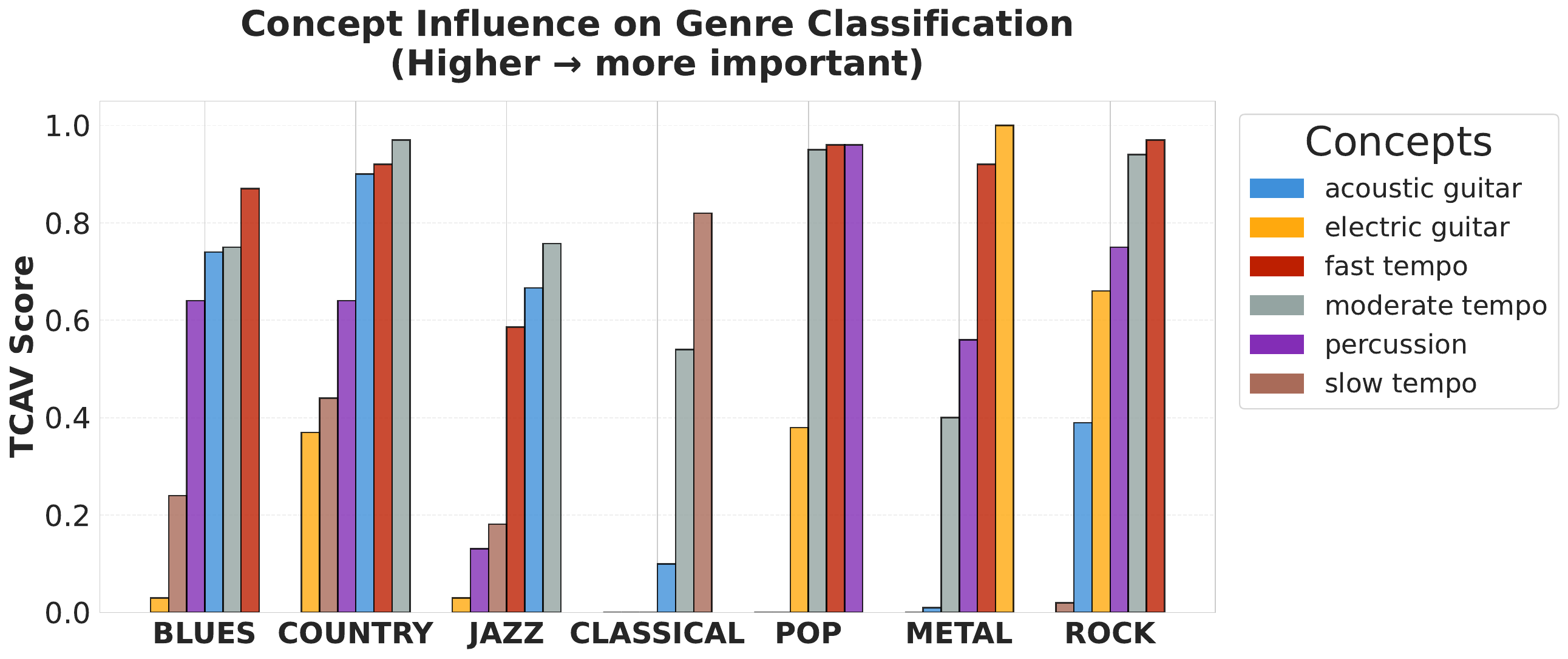}
    \caption{TCAV scores illustrating concept importance across genres. Higher scores indicate that a concept serves as a highly discriminative feature for a specific class within the model's latent space.}
    \label{fig:tcav_results}
\end{figure}

The results presented in Figure \ref{fig:tcav_results} confirm that the classifier has learned domain-appropriate musical concepts. The TCAV scores reveal clear variations in concept importance that align with musicological expectations. Notably, the importance of "slow tempo" varies significantly across genres, being highly relevant for Classical music ($\sim$0.8) while providing low relevance for genres like Metal.

Furthermore, instrumentation concepts show high class-conditional relevance; for example, "acoustic guitar" reaches high importance within the Country class ($\sim$0.9), which aligns with its characteristic sound. Conversely, "electric guitar" exhibit inverse relevance patterns, being highly important for Rock and Metal while irrelevant for Classical music.

With this analysis, we demonstrate that the classifier's internal representations are meaningfully structured around musically relevant concepts, validating the effectiveness of our distilled dataset for concept-based interpretability.

\section{Conclusions}
\label{sec:conclusion}

We present ConceptCaps, a dataset of 21k music-description pairs with validated concept labels, built through a two-stage pipeline that separates semantic modeling (VAE) from linguistic generation (fine-tuned LLM). This separation yields measurable gains in both caption quality and audio-text alignment over monolithic approaches like LP-MusicCaps or zero-shot generation. Our TCAV experiments confirm that the dataset enables meaningful concept-based probing of music classifiers. 
While extending the taxonomy to non-Western musical traditions remains a next step, we hope ConceptCaps will facilitate further research in interpretability of music models.

\subsection{Limitations}

While our distilled dataset offers significant improvements in concept clarity and alignment, several limitations remain. First, the \textbf{dependence on upstream models} means our dataset inevitably inherits the biases and artifacts of its generator models. The audio generation relies on MusicGen, which has been shown to exhibit Western-centric biases. Similarly, the Llama 3-based text generation, despite fine-tuning, may occasionally produce "hallucinated" or verbose details, which reduces audio concept alignment.

Second, the distillation process itself introduces a \textbf{selection bias}. By filtering for "high-confidence" and "coherent" attribute combinations, we potentially exclude experimental, avant-garde, or cross-cultural musical concepts that do not fit the VAE’s learned distribution of "plausibility." This results in a cleaner, but perhaps less creatively diverse, representation of the musical landscape compared to raw, noisy web data.

\subsection{Future Work}

We belive that this work introduces a strong foundation for future research in music dataset generation and interpretability. Having direct access to quality concept dataset will facilitate development of new XAI methods specifically tailored for music models, making them more transparent and trustworthy. This is especially important as music generation models become more prevalent in creative industries, raising concerns about authorship, originality, and ethical use.

While present in our dataset, \textbf{aspects related to human voice are not represented in resulting audio} due to additional requirement of providing lyrics and voice synthesis for each sample, which is out of scope for this thesis. We decided to still include such samples as they represent important musical concepts related to vocal music, singing styles, and lyrical themes.
Future work could explore integrating lyrics generation to create vocal music samples, further enriching the dataset's diversity.

Introducing \textbf{human in the loop} during the generation process could further enhance the quality of both captions and audio. For example, human reviewers could validate or refine generated captions before audio synthesis, ensuring higher semantic fidelity. Similarly, human evaluation of audio samples could help identify and filter out low-quality or misaligned generations. Incorporating human feedback loops would increase the dataset's reliability for interpretability research.

Future work could also explore \textbf{expanding the taxonomy of musical concepts} used in the VAE. By providing source code along with models used in the generation pipeline, researchers can adapt the model to include more diverse or specialized musical attributes, such as cultural genres, production techniques, or emotional nuances. This would allow the creation of tailored datasets for specific research questions or applications, further enhancing the utility of the proposed approach.

\appendix

\section*{Ethical Statement}

\paragraph{Copyright and Fair Use.}
A primary motivation for this work is to address the legal and ethical bottlenecks in music AI research \cite{barnett2025ethicsstatementsaimusic}. Standard datasets like AudioSet or valid portions of MusicCaps often rely on Western-biased, copyrighted commercial music, creating legal uncertainty for downstream model distribution. By releasing a fully synthetic, copyright-free dataset, we aim to provide a safe solution for researchers to benchmark interpretability tools without infringing on intellectual property rights. However, we acknowledge the ethical complexity of "laundering" musical concepts: while the specific audio is synthetic, the underlying generative models are trained on vast corpora of human artistry, often without consent or compensation.

\paragraph{Environmental Impact.}
The creation of this dataset involved approximately 100 GPU-hours of inference (MusicGen and Llama 3). While this one-time cost is significant, we believe the release of a reusable, high-quality dataset will reduce the need for individual researchers to run redundant, large-scale generation cycles, ultimately saving computational resources in the long term.

\section*{Acknowledgments}
% TODO deanonymize
We gratefully acknowledge Polish high-performance computing infrastructure PLGrid (HPC Center: ACK Cyfronet AGH) for providing computer facilities and support within computational grant no. PLG/2025/018397.
% Anonymized for the blind review

% \mateusz{todo}

\bibliographystyle{named}
\bibliography{refs}

\end{document}